\documentclass[twocolumn,showpacs,preprintnumbers,amsmath,amssymb,groupaddress]{revtex4}
\usepackage{bbm}
\usepackage{amsfonts}
\usepackage{amsmath}
\usepackage{graphicx}% Include figure files
\usepackage{dcolumn}% Align table columns on decimal point
\usepackage{bm}% bold math

\begin{document}

%\preprint{CREAM/MS-2004-01-PRA-GH-01}

\title{Large phase shift of spatial soliton in lead glass
by cross-phase modulation in pump-signal geometry
}% Force line breaks with \\

\author{Qian Shou, Dongwen Liu, Xiang Zhang, Wei Hu}
\affiliation{Laboratory of Photonic Information Technology, South
China Normal University, Guangzhou 510631,China}
\author{Qi
Guo} \email{guoq@scnu.edu.cn} \affiliation{Laboratory of Photonic
Information Technology, South China Normal University, Guangzhou
510631,China}

\date{\today}% It is always \today, today,

\begin{abstract}
We investigate the large phase shifts of the bi-color spatial
soliton pair in a cylindrical lead glass rod. The theoretical study
suggests a synchronous propagation of a strong pump beam and a weak
signal beam under the required initial condition. We experimentally
obtain a $\pi$ phase shift of the signal beam by changing the power
of the pump beam by about 14 mW around the soliton critical power,
which agrees qualitatively with our theoretical result. The ratio of
the phase shift rate of the signal soliton to that of the pump
soliton shows a close match to the theoretical estimation.
\end{abstract}

.

%\keywords{Use showkeys class option if keyword display desired}%
\maketitle

\section{Introduction}
In possession of infinite nonlocality in nature, lead glass is one
of the most promising candidate to serve as the propagation media
for the strongly nonlocal spatial optical soliton (SNSOS) predicted
by Snyder and Mitchell \cite{SM-Science-97}. High-order solitons
\cite{Rotschild-ol-2006}, vortex solitons {CR-prl-05}, surface
solitons \cite{BA-prl-07} have been realized in lead glass since
nonlocal nonlinear response can suppress soliton transverse
instabilities; long-range interaction between solitons
\cite{CR-np-06} as well as soliton and boundaries
\cite{Alfassi-ol-2007,Shou-ol-09} have also been carried out in lead
glass. Recently, the large phase shift of nonlocal solitons in lead
glass along with the propagation distance and the soliton power was
investigated by Shou et al \cite{Shou-ol-2011}. This is the first
theoretically and experimentally study on the nonlocal soliton phase
shift or phase modulation itself which is often regarded as a
parameter or causation in the study of the soliton interaction.
%Such solitons in lead glass exhibit a host of features that have no
%counterpart in local or nonlocal media with finite range
%nonlocality. Perhaps the most intriguing property is that

The theoretical prototype of Shou's work was addressed by Guo et al
\cite{Guo-pre-04,Xie-oqe-2004}. Based on the phenomenological
Gaussian response function, Guo et al. predicted a large phase shift
in nonlocal media which is $(w_{m}/w_{0})^{2}$ times larger than
that of the local counterpart, where $w_{m}$ and $w_{0}$ are the
characteristic length of the response function and the beam width,
respectively. In actual strongly nonlocal media, lead glass, Shou
predicted the times is much smaller, but the phase shift rate is
still more than one order larger than the result for local solitons.
They conducted the experiment in lead glass to obtain a $\pi$-phase
shift by changing the soliton power around its critical power
\cite{Shou-ol-2011}.

However, Shou's work is not practical in real optical fiber
communication system. Firstly, it is not usual to obtain phase shift
of the carrier wave by mean of self-phase-modulation. Secondly, the
soliton critical power in lead glass is much stronger than the power
of the carrier wave. In this paper a $\pi$-phase shift of the signal
SNSOS is obtained by adjusting the pump SNSOS power with the aid of
the cross modulation between the SNSOSs. Although the power of the
signal soliton is much smaller than its critical power, trapping in
the pump-soliton-induced waveguide, its phase shift can be
sensitively and linearly modulated by the power of the pump soliton.
The ratio of the phase shift of the signal beam to that of the pump
beam is determined by the ratio of their wavelengths.

\section{the solution of the coupled equations}
The system we study is described by the nonlocal nonlinear
Schrodinger equations for the two slowly varying light fields
amplitude $A_{p}$ and $A_{s}$ coupled to the steady-state heat
transfer equation. In cylindrical coordinate for $z$-axis symmetry,
the three coupled equations can be expressed \cite{Zheng-oc-09}:
\begin{subequations}\label{coupled equations}
\begin{equation}\label{pump equation}
2ik_{p}\frac{\partial A_{p}}{\partial
Z}+\frac{1}{R}\frac{\partial}{\partial R}\left (R \frac{\partial
A_{p}}{\partial R}\right )+2k_{p}^2\frac{\Delta
N_{p}}{n_{0p}}A_{p}=0,
\end{equation}
\begin{equation}\label{signal equation}
2ik_{s}\frac{\partial A_{s}}{\partial
Z}+\frac{1}{R}\frac{\partial}{\partial R}\left (R \frac{\partial
A_{s}}{\partial R}\right )+2k_{s}^2\frac{\Delta
N_{s}}{n_{0s}}A_{s}=0,
\end{equation}
\begin{equation}\label{temperature distribution equation}
\frac{1}{R}\frac{\partial}{\partial R}\left (R \frac{\partial
T}{\partial R}\right )=
-\frac{1}{\kappa}(\alpha_{p}I_{P}+\alpha_{s}I_{s}),
\end{equation}
\end{subequations}
where $I_{p,s}=\mid A_{p,s}\mid^{2}$ is the light intensity for pump
and signal beams respectively. A slight portion of the light energy
is absorbed by the glass with the absorption coefficient
$\alpha_{p,s}$ and is conducted transversely to the boundary of the
glass rod with the thermal conductivity $\kappa$. $\Delta N_{p,s}$
is the light induced refractive index which is proportional to the
change of temperature
\begin{equation}\label{thermal-refractive index equation}
\Delta N_{p,s}=\beta_{p,s}(T-T_{0}),
\end{equation}
where $T_{0}$ is the fixed temperature at the boundary and
$\beta_{p,s}$ is the thermal-optical coefficient.

In the case of $I_{s}\ll I_{p}$ we only consider the light energy
contributing from the pump beam. Simplifying Eq.(\ref{coupled
equations}) in a dimensionless form and using
Eq.(\ref{thermal-refractive index equation}), we can obtain,
\begin{subequations}\label{normalized coupled equations}
\begin{equation}\label{normalized pump equation}
2i\frac{\partial a_{p}}{\partial
z}+\frac{1}{r}\frac{\partial}{\partial r}\left (r \frac{\partial
a_{p}}{\partial r}\right )+4\Delta na_{p}= 0,
\end{equation}
\begin{equation}\label{normalized signal equation}
2i\frac{\partial a_{s}}{\partial z}+\frac{1}{\mu
r}\frac{\partial}{\partial r}\left (r \frac{\partial a_{s}}{\partial
r}\right )+4\mu\rho \Delta na_{s}= 0,
\end{equation}
\begin{equation}\label{normalized nonlinear index distribution equation}
\frac{1}{r}\frac{\partial}{\partial r}\left (r \frac{\partial \Delta
n}{\partial r}\right )=-\mid a_{p}\mid^{2},
\end{equation}
\end{subequations}
where $r=R/w_{0}$, $z=Z/(2k_{p}w_{0}^{2})$, $\Delta
n=k_{p}^{2}w_{0}^{2}\Delta N_{p}/(2n_{0p})$,
$a_{p,s}=A_{p,s}/A_{0}$,
$A_{0}^{2}=2n_{0p}\kappa/(\alpha_{p}\beta_{p}k_{p}^{2}w_{0}^{4})$,
$\mu=k_{s}/k_{p}=\lambda _{p}/\lambda _{s}$, and
$\rho=n_{0p}\beta_{s}/(n_{0s}\beta_{p})$. It is indicated by
Eq.(\ref{normalized signal equation}) and Eq.(\ref{normalized
nonlinear index distribution equation}) that the signal beam
propagates in a pump-beam-induced index distribution, say a
waveguide. Therefore the propagation behavior of the signal beam is
totally determined by that of the pump beam.

According to the Snyder's method the nonlinear index can be expanded
in a Taylor series and only the first two terms are kept. This
indicates the light-induced index acts as a parabolic waveguide in
the region of the pump beam center.
\begin{equation}\label{exploded nonlinear index}
\Delta n=\Delta n^{(0)}-r^{2}\Delta n^{(2)}.
\end{equation}
Assuming the solutions of Eq. 1 are in the forms of Gaussian
functions
\begin{subequations}\label{hypothesis solutions}
\begin{equation}\label{pump hypothesis solution}
a_{p}=\frac{\sqrt{p_{0p}}\exp[i\theta_{p}(z)]}{\sqrt{\pi}w_{p}(z)}\exp
\left [-\frac{r^{2}}{2w_{p}^{2}(z)}+ic_{p}(z)r^{2}\right ],
\end{equation}
\begin{equation}\label{signal hypothesis solution}
a_{s}=\frac{\sqrt{p_{0s}}\exp[i\theta_{s}(z)]}{\sqrt{\pi}w_{s}(z)}\exp
\left [-\frac{r^{2}}{2w_{s}^{2}(z)}+ic_{s}(z)r^{2}\right ],
\end{equation}
\end{subequations}
where $p_{0p,0s}=\int\mid
a_{p,s}(x^{'}-x_{0},y^{'})\mid^{2}dx^{'}dy^{'}$ is the normalized
light power. By integrating Eq.(\ref{normalized nonlinear index
distribution equation}) twice~\cite{Shou-ol-2011}, we obtain $\Delta
n^{(0)}=p_{0p}/(4\pi) \{\Gamma [0,r^{2}_{0}/w^{2}(z) ]+\ln
[r^{2}_{0}/w^{2}(z)]+\gamma \}$, where $\Gamma$ is the Gamma
function, $\gamma=0.58$ is Euler's constant, $r_{0}$ is the
normalized radius of the cross section of the glass rod.

$\theta_{p,s}(z)$ in Eq.(\ref{pump hypothesis solution}) and
(\ref{signal hypothesis solution}) is phase shift of the pump beam
and the signal beam, respectively. We can directly obtain the phase
shift associated with $\Delta n^{(0)}$, called the zero-order phase
shift to be
\begin{equation}\label{zero order index}
\theta_{p}^{(0)}(z)=n^{(0)}z,
\theta_{s}^{(0)}=\mu\rho\theta_{p}^{(0)}.
\end{equation}
It is obvious that the zero-order phase shifts of the pump and
signal beams are proportional to the pump power. The total phase
shift can be rewritten as, according to the two terms of the
nonlinear index designating in Eq. (\ref{exploded nonlinear index}),
\begin{equation}\label{seperated phase}
\theta_{p,s}(z)=\theta_{p,s}^{(0)}(z)+\theta_{p,s}^{(2)}(z).
\end{equation}

Following the method in Ref \cite{Guo-pre-04}, by substituting
Eq.(\ref{exploded nonlinear index})-(\ref{seperated phase}) into
Eq.(\ref{normalized coupled equations}), we can obtain the evolution
equations of the beam widths and the second-order phase shifts of
pump and signal beams,
\begin{subequations}
\begin{equation}\label{pump beam width}
\frac{d^{2}w_{p}}{dz^{2}}=\frac{4}{w^{3}_{p}}-\frac{4p_{0p}}{\pi
w_{p}},
\end{equation}
\begin{equation}\label{signal beam width}
\frac{d^{2}w_{s}}{dz^{2}}=\frac{4}{\mu^{2}w^{3}_{s}}-\frac{4\rho
p_{0p}w_{s}}{\pi w^{2}_{p}},
\end{equation}
\begin{equation}\label{second-order phase of pump beam}
\frac{d\theta_{p}^{(2)}}{dz}+\frac{2}{w_{p}^2}=0,
\end{equation}
\begin{equation}\label{second-order phase of signal beam}
\frac{d\theta_{s}^{(2)}}{dz}+\frac{2}{\mu w_{s}^2}=0.
\end{equation}
\end{subequations}
When there is not much difference between the wavelengths of the two
beams, $\rho \approx 1$. Thereby Eq.(\ref{signal beam width}) become
\begin{equation}\label{signal beam width with gamma equals 1}
\frac{d^{2}w_{s}}{dz^{2}}=\frac{4}{\mu^{2}w^{3}_{s}}-\frac{4p_{0p}w_{s}}{\pi
w^{2}_{p}}.
\end{equation}
Compare Eq.(\ref{pump beam width}) with Eq.(\ref{signal beam width
with gamma equals 1}), replace $w_{s}$ with $(1/\sqrt{\mu})w_{p}$,
these two equations turn to be equivalent mathematically. In such
case $\theta_{s}^{(2)}(z)$ and $\theta_{p}^{(2)}(z)$ have the same
evolution equations revealed in Eq. (\ref{second-order phase of pump
beam}) and Eq. (\ref{second-order phase of signal beam}). Satisfying
the initial condition $w_{0s}=(1/\sqrt{\mu})w_{0p}$, we can achieve
the synchronous propagations of the pump beam and the signal beam.

%their Rayleigh lengths are equal, this means that their
%diffraction effects are equal. so, these two beams will be broaden
%and compressed at the same time. Namely, these two beams will
%propagate synchronously. %that is $k_{s}w_{s0}^{2}=k_{p}w_{p0}^{2}$

\begin{figure}[htb]
\centerline{\includegraphics[width=8.0cm]{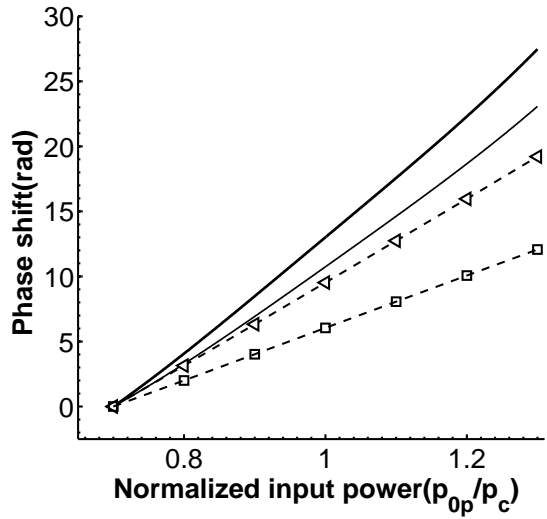}} \caption{Phase
shifts of pump beam and signal beam versus $p_{0p}/p_{c}$. Solid
thick line and thin line are respectively analytical results of pump
and signal beam in case of $r_{0}=200$ and $z=2$, which accords to
the following experiment situation. Triangle and squares are
respectively numerical results of pump and signal beam in the same
case.}\label{shuzhi jiexi bijiaotu}
\end{figure}

The critical power $p_{c}=\pi$ for the pump beam to propagate in the
form of soliton can be easily obtained based on Eq.(\ref{pump beam
width}) supposing $w_{p}(z)=1$. The signal beam also forms a soliton
in the "soliton-waveguide", though its power is far less than its
own critical power. The beam widths and the second-order phase
shifts of pump and signal beams are of the forms
\begin{equation}\label{solution of beam width}
w_{s}(z)=\frac{1}{\sqrt{\mu}}w_{p}(z)=\frac{1}{\sqrt{\mu}}[\sigma+(1-\sigma)\cos(bz)],
\end{equation}
\begin{equation}\label{solution of second-order beam phase}
\begin{split}
&\theta_{s}^{(2)}(z)=\theta_{p}^{(2)}(z)=-\frac{2}{2\sigma-1} \Bigg \{\frac{(1-\sigma)\sin(bz)}{\sigma+(1-\sigma)\cos(bz)}\\
&-\frac{2\sigma}{\sqrt{2\sigma-1}}\left [\arctan
(\sqrt{(2\sigma-1)}\tan\frac{bz}{2})+\pi \right ]\Bigg \},
\end{split}
\end{equation}
where $\sigma=\sqrt{p_{c}/p_{0p}}$, $b=2\sqrt{2}/\sigma^{2}$.
%The results above show that initial incident conditions determine the propagate behavior of signal beam.
%Basically, it is because, when signal beam power is far-weaker than the pump beam power, the pump beam imprints a waveguide for the signal beam,
%solution of signal beam corresponding to a mode of the waveguide.
%Here, in this paper we go into the details about phase shift of signal beam.
%Phase of signal beam is modulated by pump beam, when initial pump power changes around its critical power for the soliton propagation,
%$\theta_{p,s}=\theta^{(0)}_{p,s}+\theta^{(2)}_{p,s}$ will change almost linearly with it, the ratio of the two slopes is approach to $\mu$,
%which is determined by their wavelength.
Fig.\ref{shuzhi jiexi bijiaotu} show comparison between the
analytical results and numerical results, while the numerical
results is obtained by simulating Eq.(\ref{normalized pump
equation}, \ref{normalized signal equation}, \ref{normalized
nonlinear index distribution equation}). The disagreement between
the analytical result and the numerical result is attributed to the
limited expanding of the nonlinear index. Ma et al. also numerically
obtained a bigger change rate of the propagating constant with power
than the analytical result when they investigated the surface
soliton in lead glass~\cite{X.K.Ma-pra-11}.

\section{Experiment in lead glass}
\begin{figure}[htb]
\centerline{\includegraphics[width=9.0cm]{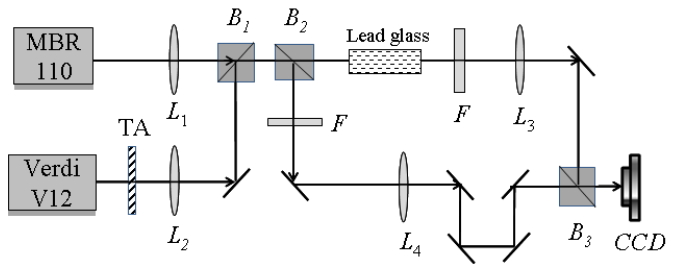}}
\caption{Experiment setup. TA is tunable attenuator; \emph{$B_{1}$}
is a polarizing cube beamsplitter, \emph{$B_{2}$} and \emph{$B_{3}$}
are non-polarizing cube beamsplitters; \emph{F} is beam filter;
\emph{$L_1$}, \emph{$L_2$}, \emph{$L_3$}, \emph{$L_4$} are
lenses.}\label{guang lu tu}
\end{figure}

We carry out the experiment in a lead glass rod using the
Mach-Zehnder interferometer technique to measure the phase shift of
the signal beam modulated by the power of the pump beam. The glass
rod has a radius of $7.5$ mm and length of $60$ mm. Its absorption
coefficient changes strongly with wavelength which is of the value
of 0.07 $\rm{cm}^{-1}$ at the pump wavelength 532 nm and 0.03
$\rm{cm}^{-1}$ at signal wavelength 790 nm. The sketch map of the
experimental setup is detailed in Fig. \ref{guang lu tu}.

A small portion of energy launched from a coherent Verdi V12 laser
contributes to act as the pump beam of orthogonal polarization. The
most portion of energy of Verdi V12 serves as the pump source for a
tunable, single-frequency ring Ti:Sapphire laser MBR 110 which
produces the signal beam of parallel polarization. The pump and
signal beams are focused by $\rm{L}_{\rm{1}}$ and $\rm{L}_{\rm{2}}$
with different beam widths of 75$\mu m$ and 90$\mu m$, respectively,
but the same positions of the focal points . $\rm{B}_{\rm{1}}$ is a
polarizing cube beamsplitter which allows signal beam transmits and
pump beam reflects into the Mach-Zehnder interferometer,
respectively. The pump beam is absorbed totally by a filter on one
arm after coaxially propagating through the lead glass with the
signal beam. On the other arm the pump beam is absorbed by another
filter just after being reflected by the non-polarizing beamsplitter
B2. The beam spot at the output side of the glass rod is imaging by
lens F3 onto the CCD where it interferences with a large-diameter
beam spot acting as a phase reference. The interference fringes move
along with the change of the pump power indicates the phase shift of
the signal beam modulated by the pump power. The phase shift of the
pump power modulated by the power of its own can be obtained in the
same way but after removing F1 and F2 and blocking the signal beam.
The rest detail of the experimental setup can be found in
Ref.\cite{Shou-ol-2011}.

\begin{figure}[htb]
\centerline{\includegraphics[width=8.0cm]{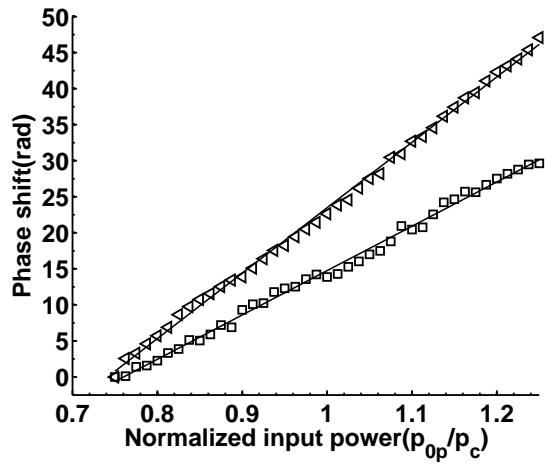}}
\caption{Experimental phase shift of pump beam and signal beam
versus $P_{0p}/P_{c}$. Pump beam wavelength is 532nm, and signal
beam wavelength 790nm; critical power Pc is measured to be 260 $mW$.
Triangle and squares are, respectively, the experimental data of
pump and signal beams. The solid curves are the linear
fits.}\label{shiyantu}
\end{figure}

By tuning the input power of pump beam from $190mW$ to $340mW$, we
obtain the phase shifts of the the pump beam or the signal beam as
function of the input power normalized by the critical power in lead
glasses. Ref. \cite{Shou-ol-2011} gives a detailed demonstration how
to obtain the phase shift from the interference fringes captured by
CCD. As shown in Fig.\ref{shiyantu}, linear fits of the data is
good, while the slopes of the two lines are $90.7(rad/P_{c})$ and
$61.5(rad/P_{c})$. Due to some reasons mentioned in
Ref.~\cite{Shou-ol-2011}, phase shift of the two beams in our
experimental is bigger than the numerical and analytical results, in
other words, the modulation sensitivity suggested by the
experimental results is about twice that predicted by the
theoretical curves in Fig.\ref{shuzhi jiexi bijiaotu}. However, The
ratio of the two slopes in the experiment is $1.47$, which is in
very good agreement with the numerical result. This means that the
ratio of the two slopes is only determined by the ratio of their
wavelength.

Fig. 4 shows the phase shifts of the soliton pair when the beam with
wavelength of 790nm serves as the pump soliton, while beam with
wavelength of 532nm serves as signal soliton. We do not know the
critical power of the pump soliton because the absorption
coefficient $\alpha$ at 780nm is so small that the critical power
for soliton propagation is too higher to be obtained. Therefore the
changing power in Fig. 4 is not normalized by critical power. The
ratio between the slopes of the two fit lines is 1.45, which also
approaches to the numerical result.
\begin{figure}[htb]
\centerline{\includegraphics[width=8.0cm]{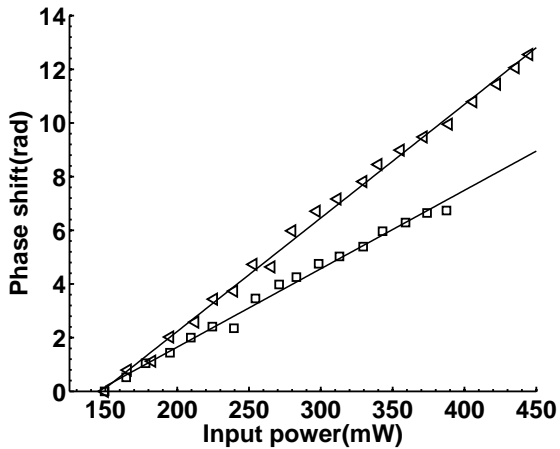}} \caption{(color
online)Experimental phase shift of pump beam and signal beam versus
input pump power. Pump beam wavelength is 790nm, and signal beam
wavelength 532nm; triangle and squares are, respectively, the
experimental data of pump and signal beams. The upper solid curves
are the linear fits.}\label{shiyantu}
\end{figure}

\section{Conclusion}
In conclusion, we investigate the large phase shifts of the bi-color
spatial soliton pair in lead glass. A 1.5$\pi$-phase shift of the
pump soliton and a $\pi$-phase shift of the signal soliton are
simultaneously obtained by tuning the power of the pump beam around
its critical power for about 14 mW via self-phase-modulation and
cross-phase-modulation, respectively. The linear modulation result
is much sensitive than the theoretical prediction, while the ratio
of the phase shift rate of the signal beam to that of the pump beam
agrees quantitatively with the result for the thermal theoretical
model. This suggests that we can obtain bigger phase shift rate by
choosing pump beam with longer wavelength. Effectively realizing the
$\pi$-phase shift is important for processing and controlling an
optical signal based on the principle of interference.

\section*{ACKNOWLEDGMENTS}
This research was supported by the National Natural Science
Foundation of China (NSFC) (Grant Nos. 60908003 and 11074080).

\end{document}